\documentclass[aps,prc,preprint,showpacs,showkeys,byrevtex]{revtex4}
\usepackage{graphicx}
\usepackage{color}
\usepackage{dcolumn}

\newcommand{\bmath}[1]{\mbox{\boldmath${#1}$}}
\newcommand{\ri}[1]{{\bf r}_{#1}}

\newcommand{\ki}[1]{{\bf k}_{#1}}
\newcommand{\bp}{{\bf p}}

\newcommand{\bsigma}{\bmath{\sigma}}
\newcommand{\btau}{\bmath{\tau}}

\newcommand{\beps}{\bmath{\epsilon}}

\newcommand{\EG}{{\textrm{e.g.}}}
\newcommand{\IE}{{\textrm{i.e.}}}
\newcommand{\EA}{{\textit{et al.}}}

\date{\today}

\begin{document}
\title{Chiral \boldmath$\mathcal{O}(Q^4)$ two-body operators for 
\boldmath$s$-wave pion photoproduction on the \boldmath$NN$ system}

\author{A. G{\aa}rdestig}\email{anders@physics.sc.edu}
\affiliation{Department of Physics and Astronomy, 
University of South Carolina, Columbia, SC 29208}

\begin{abstract}
The two-body currents for $s$-wave pion photoproduction on the $NN$ system are
derived to $\mathcal{O}(Q^4)$ in chiral perturbation theory.
For the interesting case of $^3S_1\leftrightarrow{}^1S_0$ transitions, we show
that an axial isovector two-nucleon contact term connects the short-distance 
physics of pion photoproduction to pion production and several important
electroweak reactions.
We also find that the standard chiral Lagrangian gives a $\gamma\pi\pi NN$ 
vertex that have not been explicitly mentioned in previous literature.
The corresponding Feynman rule is presented here and some processes where
it should be important are briefly discussed.
\end{abstract}

\pacs{21.45.+v, 12.39.Fe, 13.75.Cs, 25.80.Hp}
\keywords{chiral perturbation theory, pion photoproduction}

\maketitle
Photo- and electroproduction of pions on nucleons have been intensely studied 
in the framework of chiral perturbation theory ($\chi$PT).
Calculations exists for $\gamma d\to\pi^0d$~\cite{silasgd}, 
$\gamma^\ast d\to\pi^0d$~\cite{ulfgd}, 
$\gamma^{(\ast)} N\to \pi N$~\cite{emprod},
$\gamma p\to\pi^0\pi^0 p$~\cite{photoprod2}, and 
$\gamma d\to nn\pi^+$~\cite{Lensky}.
Recently, also radiative pion absorption on the deuteron has been calculated 
using $\chi$PT~\cite{GP1,GP2}.
The latter work is motivated by the crucial importance the $\pi^-d\to nn\gamma$
measurements play for the extraction of the neutron-neutron scattering length 
($a_{nn}$)~\cite{NNreview} and the corresponding need for a precise and 
accurate theoretical calculation.

The advantage of using $\chi$PT (apart from providing a tractable low-energy 
limit of QCD) is that it is an effective field theory based on an expansion in 
the ratio $\chi=Q/M_{\rm QCD}$, where $Q\sim m_\pi$ is the size of typical 
momenta and energies in the problem ($m_\pi$ is the pion mass) and 
$M_{\rm QCD}=\mathcal{O}(1~{\rm GeV})$ is the energy scale at which this EFT
breaks down.
In the following, we will assume that $M_{\rm QCD}\equiv M$ (the nucleon mass)
and that the electron charge $e<0$ also counts as the small scale $Q$.
(A detailed review of $\chi$PT can be found, \EG, in Ref.~\cite{ulfreview}.)
This expansion not only provides a prescription on how to estimate errors, but 
also suggest what needs to be done in order to improve the precision of the 
calculation.

In this brief note we derive the $\mathcal{O}(Q^4)$ two-body 
contributions to $s$-wave pion photoproduction in Coulomb gauge.
The amplitudes for neutral pion photoproduction were used in
Refs~\cite{ulfgd}, though no expressions were given.
We will find that the standard chiral Lagrangian gives an additional 
$\gamma\pi\pi NN$ vertex that have not been explicitly mentioned in 
previous literature, though it was included in the calculations of 
\cite{ulfgd,Veronique}.
We derive the corresponding Feynman rule and discuss some reactions where it 
should be important.
We will also explicitly prove that the two-body contact term needed 
in~\cite{GP2} to reduce the theoretical error in the $\pi^-d\to nn\gamma$ 
calculations is indeed the same as the one appearing in weak $pp$ fusion, 
tritium $\beta$ decay, the hep process 
($^3{\rm He} \, p\to{}^4{\rm He}\, e^+\,\nu_e$), $\nu(\bar{\nu})d$ breakup, 
and $\mu^-d\to nn\nu_\mu$, as well as in $p$-wave pion production on two 
nucleons and the chiral three-nucleon force (3NF). 
This identity can be established only if the additional Feynman rule is 
included.

The $\pi N(N)$ terms of the chiral Lagrangian relevant for the 
$\mathcal{O}(Q^4)$ amplitudes are~\cite{ulfreview,Parkhep}
\begin{eqnarray}
  \mathcal{L}^{(1)}_{\pi N} & = &  
  N^\dagger [i v \cdot D + g_{\rm A} S \cdot u] N, \\
  \mathcal{L}^{(2)}_{\pi N+\pi NN} & = & 
  N^\dagger\left[  \frac{v^\mu v^\nu-g^{\mu\nu}}{2M}D_\mu D_\nu+c_3u\cdot u 
  +\left(c_4+\frac{1}{4M}\right)[S^\mu,S^\nu]u_\mu u_\nu \right]N \nonumber \\
  & - & 2 d_1 N^\dagger S \cdot u N N^\dagger N
  +d_2\epsilon^{abc}\epsilon_{\kappa\lambda\mu\nu}v^\kappa u^{\lambda,a}
  N^\dagger S^\mu\tau^b NN^\dagger S^\nu\tau^c N,
\label{eq:chiLag}
\end{eqnarray}
where $v$ is the nucleon four-velocity, $D$ is a (chiral) covariant derivative,
$S$ is the Pauli-Lubanski spin vector, and $g_{\rm A}$ parametrizes the 
unknown short-distance physics of the nucleon's axial-current matrix element.
The low-energy constants (LECs) $c_i=\mathcal{O}(\frac{1}{M})$ parametrize pion
rescattering on one nucleon, while the $d_i=\mathcal{O}(\frac{1}{Mf_\pi^2})$ 
parametrize the two-nucleon axial isovector contact term~\footnote{This 
follows the notation adopted by Cohen~\EA~\cite{Cohen}, \IE, these $d_i$'s 
should not be confused with the ones used for third order 
$\pi N$ scattering~\cite{ulfdi}.}.
The latter have to be determined from fits to suitable two-nucleon data.
Here $u_\mu$ is an axial four-vector which contains the pion field, and when 
expanded takes the form:
\begin{equation}
  u_\mu = -\frac{\tau^a \partial_\mu \pi^a}{f_\pi} - 
  \frac{\epsilon^{3ba}V_\mu \pi^b \tau^a}{f_\pi} + 
  A_\mu + \mathcal{O}(\pi^3)
\label{eq:umu}
\end{equation}
where $V_\mu$ ($A_\mu$) is an external vector (axial) field and $f_\pi=93$~MeV 
is the pion decay constant.
Note that only the isovector part of the vector field contributes at this
order.
($u_\mu^a$ is obtained by factoring out $\tau^a$.)

\begin{figure}[ht]
\includegraphics[width=160mm]{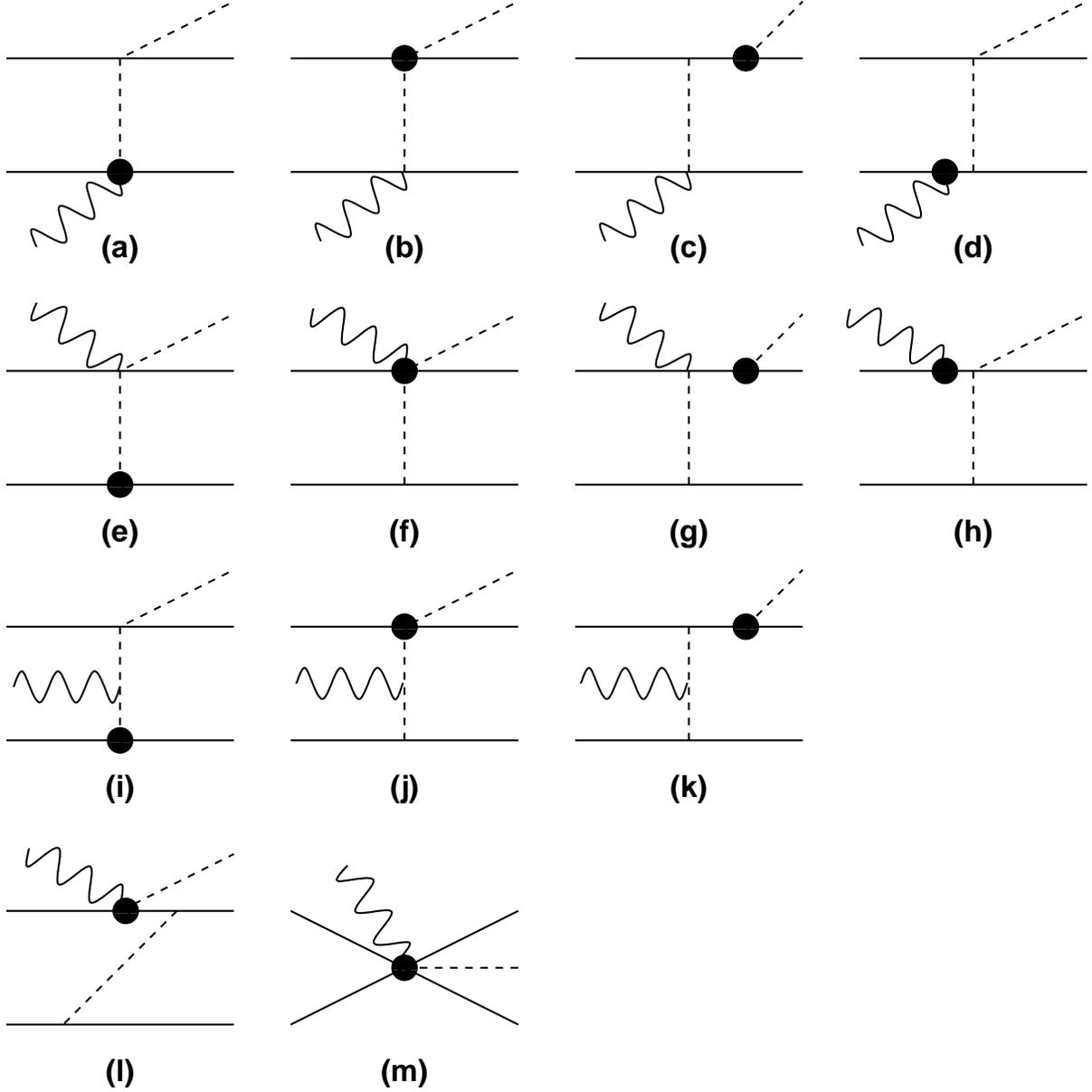}
\caption{Next-to-leading-order [$\mathcal{O}(Q^4)$] two-body diagrams for 
$s$-wave pion photo-production on two nucleons.
Only one representative vertex ordering is given for each type of diagram.
The black disc indicates an insertion from $\mathcal{L}^{(2)}_{\pi N+\pi NN}$.}
\label{fig:OQ4}
\end{figure}

At next-to-leading order [NLO or $\mathcal{O}(Q^4)$] a slew of diagrams appear,
given in Fig.~\ref{fig:OQ4}.
These diagrams are the same as the ones in Refs.~\cite{silasgd,ulfgd}, except
that here only diagrams giving $s$-wave pions are shown.
The diagrams have been grouped together according to their 
pion propagator structure.
Thus, the first row [(a)--(d)] have a Coulomb-like propagator since the energy 
of the photon has to be transferred through the exchanged pion in order to 
produce a final, real pion, while row two [(e)--(g)] has an off-shell (zero 
energy transfer) pion propagator, and row three [(i)--(k)] has one propagator 
of each kind.
The initial virtual pion in (e) and (i) is emitted in an $s$-wave (from 
$\mathcal{L}_{\pi N+\pi NN}^{(2)}$), which makes these terms proportional to 
the (in this case) very small energy of this pion.
Thus these diagrams are pushed to higher order because of kinematics and
will not be considered here.
In addition, diagram (e) will vanish in the Coulomb gauge.
The stretched pion exchange diagram (l) was included in \cite{silasgd,ulfgd}, 
where it was evaluated in time-ordered perturbation theory, but its evaluation 
depends on the way the attached wave functions are constructed, \EG, if they 
are energy-dependent or not.
For this reason, we will not derive explicit expressions for this type 
of diagram.
Finally, diagram (m) contains the two-body contact terms given by the 
LECs $d_1$ and $d_2$.
Explicit expressions for the surviving diagrams are given in the Appendix.

In principle, diagrams (c), (d), (f), (g), (h), and (k) should all have 
contact terms, \IE, they contain constant pieces that gets Fourier transformed
into $\delta^{(3)}(\ri{})$ in configuration space.
However, for $^3S_1\leftrightarrow{}^1S_0$ transitions the 
($\frac{g_{\rm A}^3}{Mf_\pi^3}$) contributions from (c) and (g) will partially 
cancel against (k), while the ($\frac{g_{\rm A}}{Mf_\pi^3}$) contributions from
(d) and (h) will be completely canceled by a term from (f).
This leaves the contact terms given by $c_3$, $c_4+\frac{1}{4M}$ [from (f)],
$d_1$, $d_2$ [from (m)], and the surviving $g_{\rm A}^3$ piece from (c), (g), 
and (k) as the only relevant ones.
The contact terms linear in $g_{\rm A}$ will always appear (for 
$^3S_1\leftrightarrow{}^1S_0$) with the combined coefficient
\begin{equation}
  \hat{d} \equiv \hat{d}_1+2\hat{d}_2+\frac{\hat{c}_3}{3}+\frac{2\hat{c}_4}{3}+
  \frac{1}{6},
\label{eq:hatd}
\end{equation}
as explained in the appendix.
The $g_{\rm A}^3$ contact term appears only in the $\pi^-d\to nn\gamma$ and
$\gamma d\to nn\pi^+$ reactions.
The reduced (dimensionless) LECs $\hat{c}_i$ and $\hat{d}_i$ are defined 
through
\begin{eqnarray}
 c_i \equiv \frac{\hat{c}_i}{M},  &&
 d_i \equiv \frac{g_{\rm A}\hat{d}_i}{Mf_\pi^2}.
\end{eqnarray}
As claimed in~\cite{GP2}, the expression (\ref{eq:hatd}) is the same 
as the one appearing in $pp$ fusion, tritium $\beta$ decay, and the hep 
reaction~\cite{Parkhep}, neutrino-induced deuteron breakup~\cite{nud}, 
muon capture on deuterium~\cite{mud}, $p$-wave pion production on two nucleons,
and the chiral 3NF~\cite{HvKM}.
Because of Fermi-Dirac statistics, the LECs $\hat{d}_i$ will always 
appear in the linear combination $\hat{d}_1+2\hat{d}_2$~\cite{Parkhep}.

The gauged $c_4$ term [derived from the Lagrangian (\ref{eq:chiLag}) and part 
of diagram (f)] has not, to the best of our knowledge, been explicitly 
mentioned in the literature before, \IE, it is missing in the Feynman rules 
derived in Ref.~\cite{ulfreview} [their Eq.~(A.31)]~\cite{Veronique}.
The Feynman rule for this term can be written (in a notation reminiscent
of~\cite{ulfreview}) as
\begin{equation}
  \left(c_4+\frac{1}{4M}\right)\frac{2ie}{f_\pi^2}
  \left[
    \left(\delta^{ab}\tau^3-\delta^{a3}\tau^b\right)
	 [S\cdot q_1,S\cdot\epsilon_\gamma]
    -\left(\delta^{ab}\tau^3-\delta^{b3}\tau^a\right)
    [S\cdot q_2,S\cdot\epsilon_\gamma]\right],
\end{equation}
where $a,1(b,2)$ annotate the incoming (outgoing) pion's isospin and 
four-momentum.
The corresponding pion rescattering diagram (f) is the only way to obtain the
$c_4$ contribution needed for $\hat{d}$ in $\pi^-d\to nn\gamma$ or 
$\gamma d\to nn\pi^+$.
In these $^3S_1\leftrightarrow{}^1S_0$ transitions, the rescattered pion 
emitted by the gauged $c_3$ and $c_4$ terms is in an $s$-wave relative to the 
nucleon pair.
This should be contrasted with $\gamma^{(\ast)} d\to\pi^0d$, where $c_4$ (but 
not $c_3$) can contribute and gives a $p$-wave pion only.
This $c_4$ term was included in the $\gamma^\ast d\to\pi^0d$ calculations 
of Refs.~\cite{ulfgd,Veronique}, but without being explicitly referred to.
It will also contribute to $\mathcal{O}(Q^2)$ (charged) double-pion 
photoproduction on a single nucleon above threshold.

Explicit expressions for the different surviving diagrams of Fig.~\ref{fig:OQ4}
are given in the appendix.
As mentioned above, there appears an extra contact term from diagrams (c), (g),
and (k) that did not appear in the other reactions mentioned.
However, it enters with a known coefficient and thus does not destroy the 
linear correlation between the Gamow-Teller matrix element and the FSI peak 
height that was used in Ref.~\cite{GP2} to reduce the theoretical error in the
extracted $a_{nn}$.
Even so, this term does shift the location of the correlation (it changes the 
FSI peak height) and is therefore necessary for an accurate extraction of 
$a_{nn}$.

In this paper we have derived the $\mathcal{O}(Q^4)$ two-body amplitudes for 
$s$-wave radiative pion capture or pion photoproduction on two nucleons.
We have also shown that the axial isovector contact term for these processes
can be parametrized by the same single constant that is used for certain
electroweak reactions (like $pp$ fusion and the hep process), $p$-wave pion 
production on two nucleons, and the chiral three-nucleon force.
This was accomplished through deriving an additional, previously not published,
Feynman rule (parametrized by $c_4$) for a photon coupling to two pions and 
one nucleon.
This vertex should play a role in $\pi^-d\to nn\gamma$~\cite{GP2,GP3} and 
$\gamma d\to nn\pi^+$, as well as in $\gamma^{(\ast)} d\to\pi^0d$~\cite{ulfgd} 
and $\gamma N\to \pi\pi N$ beyond threshold, for the latter two reactions 
resulting in $p$-wave pions.
Depending on selection rules for a particular reaction, recoil 
corrections might become important~\cite{Lensky}. 
Further work on this may be needed---together with a calculation of
$\mathcal{O}(Q^4)$ single-nucleon amplitudes and an investigation of when the 
time-ordered diagram of Fig.~\ref{fig:OQ4}(l) needs to be included.
This will also necessitate a thorough understanding of how to consistently
include $1/M$ corrections in operators and the $NN$ potentials used to derive 
wave functions.
We note that weak axial two-body currents and contact terms have been studied 
recently, from a somewhat different perspective, by Mosconi, Ricci, and 
Truhl{\'\i}k~\cite{emil}.

The results presented here are important for the calculation of 
$\pi^-d\to nn\gamma$ for two reasons.
Firstly, it supports the conclusion of Ref.~\cite{GP2} that there is a 
correlation between the short-distance physics in electroweak reactions and
$\pi^-d\to nn\gamma$.
Secondly, it gives expressions for the long-range $\mathcal{O}(Q^4)$ currents
that should give a correction to $a_{nn}$.
These currents, together with consideration of the issues raised in
Refs.~\cite{Lensky,emil}, will be included in forthcoming work on a precision 
computation of the $\pi^-d\to nn\gamma$ spectrum~\cite{GP3}.

It is a pleasure to acknowledge discussions with Fred Myhrer, 
Kuniharu Kubodera, and Daniel Phillips.
I appreciate communications from Veronique Bernard and Ulf-G. Mei{\ss}ner 
regarding their use of the Feynman rule presented here.
This work was supported in part by the National Science Foundation grant
PHY-0457014.

\appendix*
\section{Expressions for \boldmath$\mathcal{O}(Q^4)$ two-body diagrams}
We give here momentum-space expressions for the diagrams of Fig.~\ref{fig:OQ4}
for the case of $s$-wave pion photoproduction.
These formulas can also be used for $\pi^-d\to nn\gamma$ by time-reversal 
invariance.
In the equations below, we assume that $\tilde\bp=\bp''-\bp'$ and
$2{\bf P}=\bp''+\bp'$, where $\bp'(\bp'')$ is the relative momentum of the 
incoming (outgoing) nucleon pair. 
Also, $\beps_d$ and $\beps_\gamma$ are the deuteron and photon polarization
vectors, while $\ki{}$ is the photon wave vector and $\kappa_{0(1)}$ is the 
isoscalar (isovector) anomalous nucleon moment.
Everything is expressed in the overall rest frame.
When doing the replacement $(1\leftrightarrow2)$, the momenta $\tilde\bp$ and
${\bf P}$ change sign.
It is assumed that contact term contributions should not be included when 
evaluating the expressions (\ref{eq:A1})--(\ref{eq:C3}) below (\IE, that
$\delta^{(3)}(\ri{})$ terms should be dropped when performing the Fourier 
transform).
They are instead given explicitly in Eq.~(\ref{eq:CT}).

\subsection{Coulombic diagrams}
The momentum-space expressions for the first row of diagrams in 
Fig.~\ref{fig:OQ4} are
\begin{eqnarray}
  \mathcal{O}_{\rm a} & = & \frac{eg_{\rm A}}{8Mf_\pi^3}
  \frac{\bsigma_1\cdot\beps_\gamma}{(\tilde\bp-\frac{\ki{}}{2})^2}
  \left(\epsilon^{abc}\tau_1^b\tau_2^c+\epsilon^{a3b}\tau_2^b\right)m_\pi^2
  +(1\leftrightarrow2), \label{eq:A1} \\
  \mathcal{O}_{\rm b} & = & \frac{eg_{\rm A}}{16Mf_\pi^3}
  \frac{\bsigma_1\cdot\beps_\gamma}{(\tilde\bp-\frac{\ki{}}{2})^2}
  \left[ 8\epsilon^{a3b}\tau_1^b
    \left(4\hat{c}_1-2\hat{c}_2-2\hat{c}_3+\frac{g_{\rm A}^2}{4}\right)m_\pi^2
    \right. \nonumber \\
    && \left. 
    -i\left(\delta^{a3}\btau_1\cdot\btau_2-\tau_1^a\tau_2^3\right)
    \left(2{\bf P}+\frac{\ki{}}{2}\right)\cdot
    \left(\tilde\bp-\frac{\ki{}}{2}\right) 
    \right]+(1\leftrightarrow2), \\
  \mathcal{O}_{\rm c} & = & \frac{eg_{\rm A}^3}{8Mf_\pi^3}
  \frac{\bsigma_1\cdot\beps_\gamma}{(\tilde\bp-\frac{\ki{}}{2})^2}
 \left\{\epsilon^{a3b}\tau_1^b\left[
    i\bsigma_2\cdot\left(\tilde\bp-\frac{\ki{}}{2}\right)
    \times\left(2{\bf P}+\frac{\ki{}}{2}\right)\right] \right. \nonumber \\
    && \left. +i(\delta^{a3}\btau_1\cdot\btau_2-\tau_1^a\tau_2^3)
  \left(2{\bf P}+\frac{\ki{}}{2}\right)\cdot
  \left(\tilde\bp-\frac{\ki{}}{2}\right)^{\rule{0mm}{1ex}}
  \right\}+(1\leftrightarrow2), \\
  \mathcal{O}_{\rm d} & = & \frac{eg_{\rm A}}{8Mf_\pi^3}
  \frac{1}{(\tilde\bp-\frac{\ki{}}{2})^2}\left\{
  -\left(\epsilon^{abc}\tau_1^b\tau_2^c+\epsilon^{a3b}\tau_2^b\right)
  \bsigma_1\cdot\left(\tilde\bp-\frac{\ki{}}{2}\right)
  \beps_\gamma\cdot\tilde\bp \right. \nonumber \\
  && -\left(\delta^{a3}\btau_1\cdot\btau_2-\tau_1^a\tau_2^3\right)
  \left[(1+\kappa_1)\tilde\bp\cdot(\beps_\gamma\times\ki{})
    +i\bsigma_1\cdot\left(\tilde\bp-\frac{\ki{}}{2}\right)
  \beps_\gamma\cdot2{\bf P}\right]
  \nonumber \\
  && +\left.\left[(1+\kappa_0)\epsilon^{abc}\tau_1^b\tau_2^c+
    (1+\kappa_1)\epsilon^{a3b}\tau_2^b\right]
  \left[\bsigma_1\cdot\beps_\gamma\,\ki{}\cdot
    \left(\tilde\bp-\frac{\ki{}}{2}\right)-
    \bsigma_1\cdot\ki{}\,\beps_\gamma\cdot\tilde\bp\right]^{\rule{0mm}{1ex}}
  \right\} \nonumber \\ &&+(1\leftrightarrow2). 
\end{eqnarray}

\subsection{Non-Coulombic diagrams}
The momentum-space expressions for the second row of diagrams in 
Fig.~\ref{fig:OQ4} are
\begin{eqnarray}
  \mathcal{O}_{\rm e} & = & 0, \\
  \mathcal{O}_{\rm f} & = & \frac{eg_{\rm A}}{16Mf_\pi^3}
  \frac{\bsigma_1\cdot(\tilde\bp+\frac{\ki{}}{2})}
       {(\tilde\bp+\frac{\ki{}}{2})^2+m_\pi^2}
  \left[4\left(
    \delta^{a3}\btau_1\cdot\btau_2+\tau_1^3\tau_2^a-2\tau_1^a\tau_2^3\right)
    \textcolor{white}{\frac{1}{4}}\textcolor{white}{\frac{\ki{}}{2}}\right. 
    \nonumber \\ &&
    \times\left[(1+\kappa_1)\bsigma_2\cdot(\beps_\gamma\times\ki{})
      -2i\beps_\gamma\cdot{\bf P}\right] \nonumber \\
    &&+\left[\epsilon^{abc}\tau_1^b\tau_2^c
      +\left(16\hat{c}_3-1\right)\epsilon^{a3b}\tau_1^b\right]
    \beps_\gamma\cdot\tilde\bp \nonumber \\
    &&\left.
    +2\left(4\hat{c}_4+1\right)
    \left(\tau_1^3\tau_2^a-\tau_1^a\tau_2^3\right)
    \bsigma_2\cdot\left(\tilde\bp+\frac{\ki{}}{2}\right)\times\beps_\gamma
    \right]+(1\leftrightarrow2), \\
  \mathcal{O}_{\rm g} & = & \frac{eg_{\rm A}^3}{8Mf_\pi^3}
  \frac{\bsigma_1\cdot(\tilde\bp+\frac{\ki{}}{2})}
       {(\tilde\bp+\frac{\ki{}}{2})^2+m_\pi^2}
  \left\{
  -\epsilon^{a3b}\tau_1^b\left[\tilde\bp\cdot\beps_\gamma+
    i\bsigma_2\cdot\left(2{\bf P}+\frac{\ki{}}{2}\right)
    \times\beps_\gamma\right]
  \right. \nonumber \\
  && \left.+i\left(\delta^{a3}\btau_1\cdot\btau_2-\tau_1^a\tau_2^3\right)
  \left[2{\bf P}\cdot\beps_\gamma
    +i\bsigma_2\cdot\left(\tilde\bp-\frac{\ki{}}{2}\right)
    \times\beps_\gamma\right]\right\}+(1\leftrightarrow2), \\
  \mathcal{O}_{\rm h} & = & \frac{eg_{\rm A}}{16Mf_\pi^3}
  \frac{\bsigma_1\cdot(\tilde\bp+\frac{\ki{}}{2})}
       {(\tilde\bp+\frac{\ki{}}{2})^2+m_\pi^2}
  \left[ 
  \left(\epsilon^{abc}\tau_1^b\tau_2^c-\epsilon^{a3b}\tau_1^b\right)
  \beps_\gamma\cdot\tilde\bp\right. \nonumber \\
  && \left.+\left(\delta^{a3}\btau_1\cdot\btau_2-\tau_1^3\tau_2^a\right)
       [(1+\kappa_1)\bsigma_2\cdot(\beps_\gamma\times\ki{})
	 -i2{\bf P}\cdot\beps_\gamma
       ]\right]+(1\leftrightarrow2).
\end{eqnarray}

\subsection{Double-propagator diagrams}
The momentum-space expressions for the third row of diagrams in 
Fig.~\ref{fig:OQ4} are
\begin{eqnarray}
  \mathcal{O}_{\rm i} & = & 0, \\
  \mathcal{O}_{\rm j} & = & \frac{eg_{\rm A}}{8Mf_\pi^3}
  \frac{\beps_\gamma\cdot\tilde\bp\bsigma_1\cdot(\tilde\bp+\frac{\ki{}}{2})}
       {(\tilde\bp-\frac{\ki{}}{2})^2[(\tilde\bp+\frac{\ki{}}{2})^2+m_\pi^2]}
       \left[ -8\epsilon^{a3b}\tau_1^b
	 \left(4\hat{c}_1-2\hat{c}_2-2\hat{c}_3+\frac{g_{\rm A}^2}{4}\right)
	 m_\pi^2 \right. \nonumber \\
 && \left. 
   +i\left(\delta^{a3}\btau_1\cdot\btau_2-\tau_1^a\tau_2^3\right)
   \left(2{\bf P}+\frac{\ki{}}{2}\right)\cdot
   \left(\tilde\bp-\frac{\ki{}}{2}\right)
       \right]+(1\leftrightarrow2), \\
  \mathcal{O}_{\rm k} & = & \frac{eg_{\rm A}^3}{4Mf_\pi^3}
  \frac{\beps_\gamma\cdot\tilde\bp\bsigma_1\cdot(\tilde\bp+\frac{\ki{}}{2})}
       {(\tilde\bp-\frac{\ki{}}{2})^2[(\tilde\bp+\frac{\ki{}}{2})^2+m_\pi^2]}
   \left\{\epsilon^{a3b}\tau_1^b\left[\left(\tilde\bp-\frac{\ki{}}{2}\right)^2
     \right.\right.\nonumber \\
     &&\left.+i\bsigma_2\cdot\left(2{\bf P}+\frac{\ki{}}{2}\right)
     \times\left(\tilde\bp-\frac{\ki{}}{2}\right)\right] \nonumber \\
   && \left. -i\left(\delta^{a3}\btau_1\cdot\btau_2-\tau_1^a\tau_2^3\right)
   \left(2{\bf P}+\frac{\ki{}}{2}\right)\cdot
   \left(\tilde\bp-\frac{\ki{}}{2}\right)^{\rule{0mm}{1ex}}
   \right\}+(1\leftrightarrow2).
\label{eq:C3}
\end{eqnarray}

\subsection{Contact term}
The contact term that contributes to $^3S_1\leftrightarrow{}^1S_0$ transitions
at $\mathcal{O}(Q^4)$ is given by
\begin{eqnarray}
  \mathcal{O}_{\rm CT} & = & \frac{eg_{\rm A}}{Mf_\pi^3}
  \left[
    \left(\hat{d}_1+\frac{\hat{c}_3}{3}-\frac{g_{\rm A}^2}{12}\right)
    \epsilon^{a3b}
    (\tau_1^b\bsigma_1\cdot\beps_\gamma+\tau_2^b\bsigma_2\cdot\beps_\gamma)
    \right.\nonumber \\
    &&\left.+\left(\hat{d}_2+\frac{\hat{c}_4}{3}+\frac{1}{12}
    -\frac{g_{\rm A}^2}{24}\right)
    (\tau_1^a\tau_2^3-\tau_1^3\tau_2^a)
    \beps_\gamma\cdot(\bsigma_1\times\bsigma_2)
  \right].
\label{eq:CT}
\end{eqnarray}
Obviously, this contact term can not contribute to processes involving a 
neutral pion.
When put into a $^1S_0\to{}^3S_1$ matrix element, specialized to
$\pi^-d\to nn\gamma$ (or rather $\gamma nn\to d\pi^-$), and summed over nucleon
spins and isospins, this operator reduces to
\begin{equation}
  \frac{2ieg_{\rm A}}{Mf_\pi^3}\left(\hat{d}-\frac{g_{\rm A}^2}{6}\right)
  \beps_d^\dagger\cdot\beps_\gamma,
\end{equation}
where $\hat{d}$ is given by Eq.~(\ref{eq:hatd}).
The $g_{\rm A}^2$ term is the extra piece coming from diagrams (c), (g), and 
(k).

\bibliographystyle{apsrev}

\begin{thebibliography}{99}

\bibitem{silasgd}
S.~R.~Beane, C.~Y.~Lee, and U.~van~Kolck,
Phys.\ Rev.\ C {\bf 52}, 2914 (1995);
S.~R.~Beane, V.~Bernard, T.-S.~H.~Lee, U.-G.~Mei{\ss}ner, and U.~van~Kolck,
Nucl.\ Phys.\ {\bf A618}, 381 (1997).

\bibitem{ulfgd}
H.~Krebs, V.~Bernard, and U.-G.~Mei{\ss}ner,
Nucl.\ Phys.\ A {\bf 713}, 405 (2003);
Eur.\ Phys.\ J. A {\bf 22}, 503 (2004).

\bibitem{emprod}
V.~Bernard, N.~Kaiser, and U.-G.~Mei{\ss}ner,
Nucl.\ Phys.\ A {\bf 607}, 379 (1996);
H.~W.~Fearing, T.~R.~Hemmert, R.~Lewis, and C.~Unkmeir, 
Phys.\ Rev.\ C {\bf 62}, 054006 (2000);
V.~Bernard, N.~Kaiser, and U.-G.~Mei{\ss}ner, 
Eur.\ Phys.\ J. A {\bf 11}, 209 (2001).

\bibitem{photoprod2}
V.~Bernard, N.~Kaiser, and U.-G.~Mei{\ss}ner,
Phys.\ Lett.\ B {\bf 382}, 19 (1996).

\bibitem{Lensky}
V.~Lensky {\it et al.},
Eur.\ Phys.\ J.\ A {\bf 26}, 107 (2005).

\bibitem{GP1}
A.~G{\aa}rdestig and D.~R.~Phillips,
Phys.\ Rev.\ C {\bf 73}, 014002 (2006).

\bibitem{GP2}
A.~G{\aa}rdestig and D.~R.~Phillips,
[arXiv.org/abs/nucl-th/0603045].

\bibitem{NNreview}
R.~Machleidt and I.~Slaus,
J. Phys.\ G: Nucl.\ and Part.\ Phys.\ {\bf 27}, R69 (2001).

\bibitem{ulfreview}
V.~Bernard, N.~Kaiser, and U.-G.~Mei{\ss}ner,
Int.\ J. Mod.\ Phys.\ E {\bf 4}, 193 (1995).

\bibitem{Veronique}
V.~Bernard and Ulf-G.~Mei{\ss}ner, private communication.

\bibitem{Parkhep}
T.-S.~Park \EA,
Phys.\ Rev.\ C {\bf 67}, 055206 (2003).

\bibitem{Cohen}
T.~D.~Cohen, J.~L.~Friar, G.~A.~Miller, and U.~van~Kolck,
Phys.\ Rev.\ C {\bf 53}, 2661 (1996).

\bibitem{ulfdi}
Nadia Fettes, Ulf-G. Mei{\ss}ner, and Sven Steininger, 
Nucl.\ Phys.\ A {\bf 640}, 199 (1998).

\bibitem{nud}
S.~Ando \EA,
Phys.\ Lett.\ B {\bf 555}, 49 (2003).

\bibitem{mud}
S.~Ando \EA,
Phys.\ Lett.\ B {\bf 533}, 25 (2002).

\bibitem{HvKM}
C.~Hanhart, U.~van~Kolck, and G.~A.~Miller,
Phys.\ Rev.\ Lett.\ {\bf 85}, 2905 (2000).

\bibitem{GP3}
A.~G{\aa}rdestig and D.~R. Phillips, in progress (2006).

\bibitem{emil}
B.~Mosconi, P.~Ricci, and E.~Truhl{\'\i}k,
Eur.\ Phys.\ J. A {\bf 25}, 283 (2005);
[arXiv.org/abs/nucl-th/0212042].

\end{thebibliography}

\end{document}